# High Luminance Organic Light-Emitting Diodes with Efficient Multi-Walled Carbon Nanotube Hole Injectors

Shengwei Shi and S. Ravi P. Silva*

Nanoelectronics Center, Advanced Technology Institute, University of Surrey, Guildford, GU2 7XH, Surrey, United Kingdom

* Corresponding author: Fax: +44 (0)1483 686081. E-mail addresses: s.silva@surrey.ac.uk (S. Ravi P. Silva)

ABSTRACT:

ABSTRACT:

We report high luminance organic light-emitting diodes by use of acid functionalized multi-walled carbon nanotube (o-MWCNTs) as efficient hole injector electrodes with a simple and solution processable device structure. At only 10 V, the luminance can reach nearly 50,000 cd/m$^2$ with an external quantum efficiency over 2% and a current efficiency greater than 21 cd/A. The investigation of hole-only devices shows that the mechanism for hole injection is changed from injection limited to bulk limited because of the higher effective work function of the anode modified by the o-MWCNTs. We expect the enhancement of the local electric field brought about by both the dielectric inhomogeneities within the o-MWCNT containing anode and the high aspect ratio carbon nanotubes, improves hole injection from the anode to organic active layer at much lower applied voltage.

1. Introduction

Organic light-emitting diodes (OLEDs) are enjoying enormous interest due to their potential applications in flat panel displays and solid state lighting, with significant advances being made in the last two decades [1,2]. It is well known that the performance of OLEDs is largely dominated by charge injection from both electrodes [3]. To date indium tin oxide (ITO) has generally been used as the anode of choice in OLEDs because of its high transparency and appropriately large conductivity. However, the surface of ITO is chemically and physically ill defined, which may degrade the performance of the hole-injecting electrode in OLEDs and other applications with time. Many methods have been successfully used to modify the surface of ITO, such as ultraviolet ozone cleaning and oxygen plasma exposure, which are considered to increase the work function of ITO and thus enhance hole injection [4,5]. Moreover, direct hole-injection from ITO in most organic materials is inefficient due to the energy level mismatch at the interface. High operating voltages are needed to overcome the injection barrier, resulting in reduced efficiency. Therefore, various hole-injection layers have been incorporated at the ITO-organic interface to improve hole injection with a view of energy level matching, such as in the case of copper phthalocyanine (CuPc) [6], 4, 4', 4"-tris-(3-methylphenylphenylamino)triphenylamine (m-MTDATA) [7], polyethylene dioxythiophene:polystyrene sulfonate (PEDOT:PSS) [8], and transition metal-oxides [9,10]. It is well known that interposing a hole injection layer of CuPc significantly improves the device lifetime, but strong absorption in the visible region, particularly in the red color, decreases the electroluminescent (EL) efficiency [11]. PEDOT:PSS is another widely used hole injection layer in polymer light-emitting diodes (PLEDs), although the aqueous PEDOT:PSS dispersion can cause degradation due to its acidic nature and the presence of moisture, leading to reduced device lifetime [12,13]. The introduction of PEDOT:PSS also decreases the transmittance of the ITO substrate, which is not helpful in light management [14]. A further alternative is offered by carbon nanotubes (CNTs), including single-walled (SW) and multi-walled (MW). This is due to their remarkable properties, including high electrical conductivity, mechanical strength, excellent chemical stability and tunable work function (4.5–5.1 eV) [15]. Another aspect less considered is their excellent thermal conductivity, particularly when operating under high bias conditions.

CNTs have recently emerged as versatile material for applications in organic electronics including OLEDs and organic photovoltaics (OPVs) [15-18]. Currently, a large quantity of effort has been expended in the field of applications of SWCNTs to OLEDs because of their higher conductivity [15,16,19-21], but less so to MWCNTs. MWCNTs are easier to process than SWCNTs, because they are less prone to forming tight bundles and are significantly more inexpensive. Also, because of the large number of concentric cylindrical graphitic tubes, MWCNTs are perhaps even more suitable for achieving charge transfer and charge transport due to the predicable HOMO-LUMO energy levels and metallic conductivity [22]. Generally, MWCNTs are employed either in the form of a composite with conjugated polymers or as plain sheets in OLEDs and they have been employed as hole-injecting electrodes, charge transport layers, etc [23-29]. But, in each of these cases, high operating voltages have been applied in order to obtain moderate luminance and output efficiency. To our knowledge, the device performance is still not satisfactory with the introduction of MWCNTs in OLEDs.

In fact, there are several papers on the mechanism of charge injection from MWCNTs into OLEDs [23-27] and OPVs [30-34], but they are mainly used in composites blended with PEDOT:PSS or polymer active layers by which a sharp increase of electrical conductivity can be obtained even for very low concentration of nanotubes [23,26,27]. For example, improved luminance intensity and a decrease in the turn-on voltage in the OLEDs were obtained with the nanocomposites of PEDOT:PSS and MWCNTs as hole injection layers, with the highest reported luminance increased from 4,000 cd/$m^2$ at 20 V to 6,800 cd/$m^2$ at 15 V. The mechanism was attributed to the resistance decrease of the nanocomposites and the improvement of hole injection ability of the PEDOT:PSS by the MWCNT fillings [24,25]. MWCNTs were also reported to act as exciton dissociation centres in OPVs to get efficient charge transfer, and provide ballistic pathways for the electrons with the subsequent increase in the effective carrier mobility in the active layer, created by a percolation network of nanotubes [31,32]. But the introduction of only MWCNTs as a buffer layer at the anode or at the cathode is still novel, with no reported cases with the very high luminance [35], in particular at low turn-on voltages in the literature for OLEDs. The MWCNT layer was used as the hole-collecting electrode in photovoltaic devices, but this was because of its relatively high conductivity and high work function [36], as well as greater transparency in the near-infrared as compared to PEDOT:PSS films [37]. MWCNTs has been used as a cathode buffer layer for OLEDs to demonstrate increased electron injection and luminance characteristics, which is due to the enhancement of the local electric field and the reduction of the LUMO of the organic material [38].

Here, we report a high luminance OLED with MWCNTs as efficient hole injectors for ITO substrates. The MWCNTs are acid functionalized (o-MWCNTs) using a general chemical method, and are prepared for use in very low concentrations of 0.01 mg/ml with the solvent of an ethanol-deionized (DI) water in a volume ratio of 50:50. Raman spectroscopy and transmission electron microscopy (TEM) have allowed us to conclude that the surface damage is minimal (see supporting information). All devices are fabricated on pre-patterned ITO with N,N'-bis(3-methylphenyl)-N,N'-diphenyl-1,1'-biphenyl-4,4'-diamine (TPD) as the hole transport layer and Tris(8-hydroxyquinolinato)aluminum (Alq$_3$) as the electron transport and emission layer. We demonstrate that the o-MWCNT film is uniform with the nanotube distribution characterized by atomic force microscopy (AFM). Device studies show that the hole injection is enhanced by the use of o-MWCNT layers, and we show that the hole injection at this interface is achieved at much lower applied voltages due to the enhancement of the local field at the tip of the nanotubes [39].

2. Experimental

2.1 Acid functionalization of MWCNTs

250mg MWCNTs were added to 15ml of a 3:1 mixture of concentrated Sulphuric:Nitric acid in a round bottom flask and mixed for 10 minutes in an ultrasonic bath. After sonication the mixture was refluxed over an oil bath at 130 $^{\circ}$C for 1 hour. After allowing the mixture to cool, it was carefully diluted to 80 ml using MilliQ de-ionized (DI) water and transferred into two 50 ml centrifuge tubes and centrifuged at 8500 rpm for 20 min. The supernatant (brown liquid) was removed using a vacuum filtration process leaving a black precipitate which was then diluted with MilliQ DI water and the precipitate suspended using a vortex mixer. This process was repeated twice to remove most of the concentrated acid used in the reaction, leaving a black liquid with no obvious phases. Centrifuging was continued for 20 min at a time and the black liquid decanted into clean centrifuge tubes until no obvious precipitate remained.

This solution is filtered over a 0.1 µm polycarbonate membrane filter, whilst washing with DI water until approximately pH7 is reached and finally washed with absolute ethanol. This final product of o-MWCNTs is dried in a vacuum dessicator and the dry weight obtained. The dry o-MWCNTs were redispersed into 50% Ethanol/DI water at a concentration of 10 mg/ml over a period of 4 days on an orbital mixer, and low concentrations (0.01 mg/ml in this experiment) were diluted for use.

2.2 Device fabrication

The o-MWCNTs solution had ultrasonic treatment for 10 minutes before spin-coating on oxygen-plasma precleaned ITO substrates. The spin-coating was done at 1000 rpm for 1 minute, then films were dried on a hot plate at $160^{\circ}C$ for 5 minutes to remove any solvents, and finally ITO substrates were loaded into the evaporator chamber for device fabrication. The device structure is anode/TPD (x nm)/Alq$_3$ (y nm)/LiF (1.0 nm)/Al (150 nm), in which the anode means ITO or ITO/o-MWCNTs, and x or y indicates the thickness variables for optimization. For hole-only devices, the structure is anode/TPD (120 nm)/Al (100 nm). All evaporations were conducted under a vacuum of $4 \times 10^{-6}$ Torr. The deposition rates were controlled by a quartz oscillating thickness monitor. The current density–voltage-luminance characteristics were measured using Keithley (2400 and 2100) sources with a calibrated silicon photodiode. The morphology was characterized by atomic force microscopy (AFM) (Veeco Dimension 3100) in tapping mode. The active area of the devices was 3.9 mm$^2$. All measurements were performed without encapsulation in an ambient atmosphere.

3. Results and discussion

With the o-MWCNTs/ITO/glass assembly as the anode of OLEDs, the optical transmittance is not significantly reduced from the reference, which is shown in Figure 1. The transmittances for both substrates are near-identical, especially in the visible region, and thereby not compromising the light emission of the OLEDs.

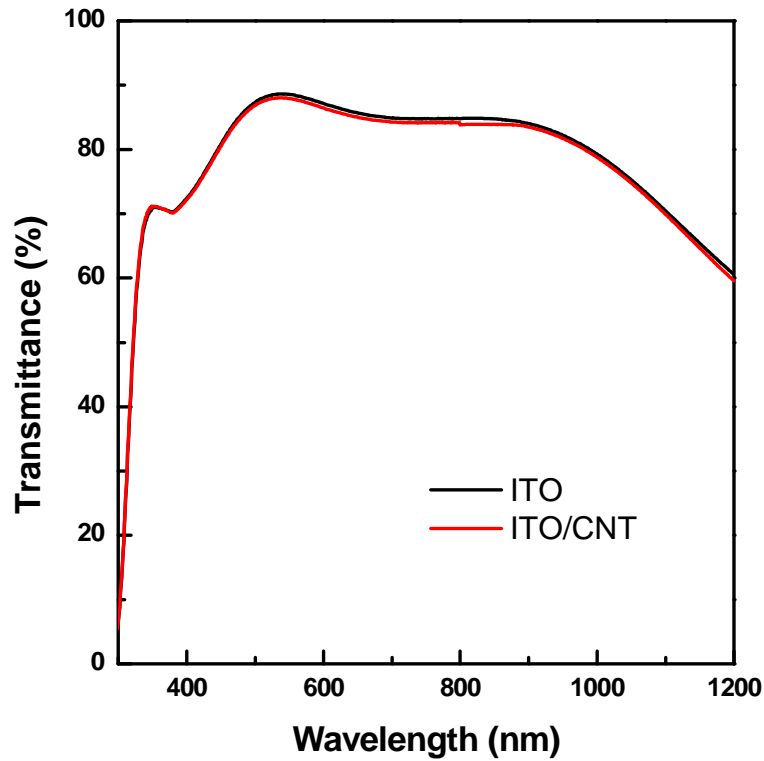

**Figure 1.** Transmittance for pure ITO (black) and modified ITO (red) by o-MWCNTs.

In terms of optimized concentrations of o-MWCNTs, we find the low concentration (0.01 mg/ml) reported here is the best. High concentrations of o-MWCNTs increase current conduction and thus device current, but they also decrease optical transmittance through the ITO and thereby device performance. Indeed, during the experiments we find that higher concentrations and enhanced coverage of the CNTs tend to destroy the diode structures, which we believe is due to electrode short circuits through stray CNTs standing on the surface. Based on the hole injection layer of o-MWCNTs, we have also optimized the thickness for TPD and Alq$_3$ according to whole device performance. Our results here are all based on the optimized thickness combination of TPD/Alq$_3$ (20 nm/60 nm), except those stated.

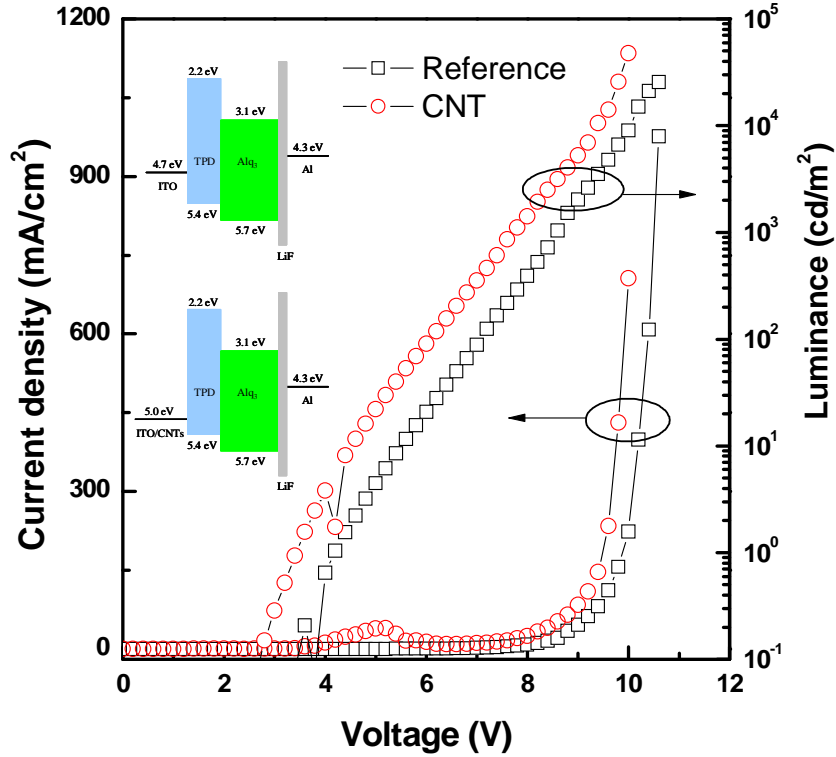

**Figure 2.** Current density-Voltage-Luminance characteristics for OLEDs with pure (black square) and modified ITO (red circle) by o-MWCNTs. Inserted are the schematic diagrams for energy levels in the experiment for (top) reference device, and (down) target device with o-MWCNT modified ITO.

Figure 2 shows current density-voltage-luminance characteristics for reference and target devices with modified ITO substrates. Our target device shows a much lower turn-on voltage (defined as the voltage required to obtain a luminance of 1 cd/m$^2$) of 3.4 V compared to that of 4.2 V for reference devices. Only at 10 V does the luminance reach the peak of 47,933 cd/m$^2$ for the target device, while the luminescence is only 9,046 cd/m$^2$ at 10 V for the reference device and its peak luminance (25,534 cd/m$^2$) appears at 10.6 V. The measurements show more than a four times increase in luminance at 10 V for the optimized device. In addition, the device current is improved with the introduction of o-MWCNTs as the hole injection layer on ITO. Since both devices have the same structure on the cathode side, we deduce that the improved hole injection results in higher device current density. To our knowledge, there are only three reports on high luminance from comparable undoped $Alq_3$ fluorescent devices [40-42]. The very recent highest value reported is 127,600 cd/m$^2$ with all carrier Ohmic-contacts by use of a complicated p-doping technology and fullerene ($C_{60}$) contact with a LiF/Al cathode [40]. The second report shows ~70,000 cd/m$^2$ luminance with 2,9-Dimethyl-4,7-diphenyl-1,10-phenanthroline (BCP) in direct contact with LiF/Al cathode, which increases to ~90,000 cd/m$^2$ at a relatively high voltage of 15.5 V with post-packaging annealing. The corresponding 10 V luminance for this device is less than 30,000 cd/m$^2$ [41]. The third report has a luminescence of 54,000 cd/m$^2$ with optimized thicknesses of Ba/Al bilayer cathodes, but the Ba is very sensitive to the environment [42]. For all of the above reported records, the devices are measured with encapsulation and N, N'-bis(l-naphthyl)-N, N'-diphenyl-1,1'-biphenyl-4,4'-diamine (NPB) as the hole transport layer. Compared to the reported best results, we obtain very high luminance of ~50,000 cd/m$^2$ at a low voltage (10 V) using simple device fabrication processes and without device encapsulation or post processing. Importantly, this is the best reported luminance to date, using MWCNT hole injectors, which may lead to solution processable large area hole injection electrodes for future OLEDs.

    Although in principle the hole carriers have a higher mobility than electrons, the hole injection to the reference device has a higher barrier than the electron with LiF/Al cathode (inserted in Fig2). In the target device, due to the relatively small "effective" hole barrier of ~0.4 eV compared to that of the ITO reference electrodes (0.7 eV), more holes are injected towards the TPD/$Alq_3$ interface, and the hole space charge improves band bending at TPD/$Alq_3$ interface which lowers the electron barrier, that enables more electrons to reach the emission interface and recombine with the holes. More holes and electrons will recombine to give more radiative emission, which improves the EL luminance and efficiency. The electrical measurement for hole-only devices in Fig 4 also provides evidence of the improvement in the hole injection level by the introduction of the o-MWCNTs on to the ITO surface. In addition, if the improvement we observe with hole injection works with undoped device performance, it is expected that in the case of the doped materials similar improvements will also result.

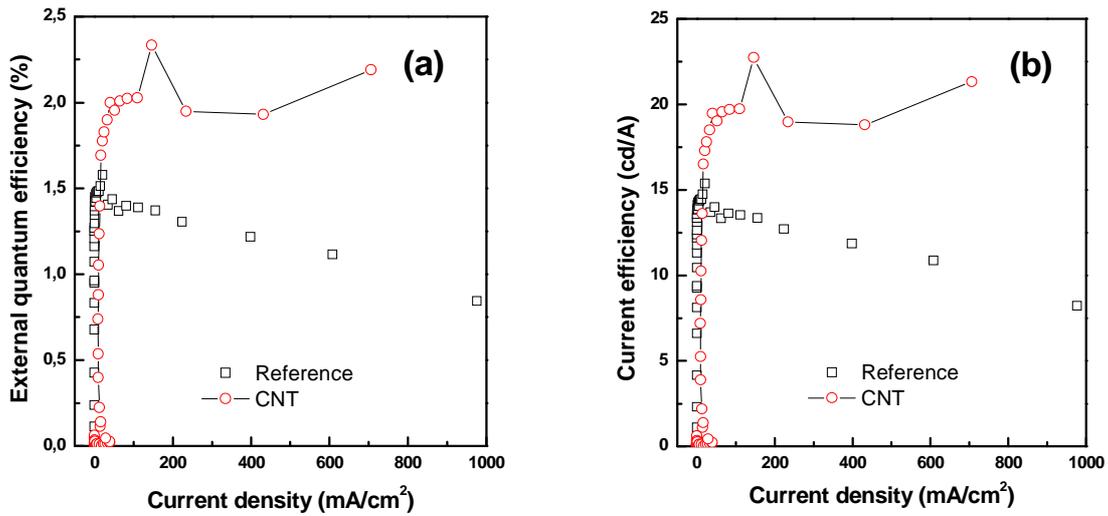

**Figure 3 (a).** External quantum efficiency (EQE)- Current density and **(b)** Current efficiency- Current density characteristics for OLEDs with pure (black square) and modified ITO (red circle) by o-MWCNT.

The external quantum efficiency (EQE)-current density characteristics for our reference and target OLEDs is shown in Figure 3a. The maximum EQE is improved from 1.57 % to 2.34 % by the introduction of o-MWCNTs on ITO, and there is a 50 % enhancement. Even at a high current density of 700 mA/cm$^2$, the target device still produces a high EQE of 2.2 %. It appears that the EQE of the target device is independent of the drive voltage and over 1.9 % in a wide range of current density from 40 to 700 mA/cm$^2$. As the EQE is proportional to the ratio of the light output to the total injected current, the voltage independent EQE in the target device is postulated to be due to a balanced injection of holes and electrons, which help enhance efficiency and stable light output from the OLEDs. In addition, a voltage independent EQE shows that the current conduction is not injection limited but bulk limited, which may be explained by the lower effective barrier height of ~0.4 eV at ITO/o-MWCNTs interface (inserted in Figure 2) and the space-charge-limited current (SCLC) behavior in the current density-voltage (J-V) relation (Figure 4). While the voltage dependent EQE in the reference device may indicate that the current conduction is injection limited (hole in this case), which results in the power-law dependence in the J-V relation (Figure 4), because of the larger effective barrier height of ~0.7 eV at the ITO/TPD interface (insert in Figure 2).

| Anode | Turn-on voltage (V) | Luminance (cd/m$^2$) | EQE (%) | Current efficiency (cd/A) | Power efficiency (lm/w) |
|---|---|---|---|---|---|
| ITO | 4.2 | 25,534 | 1.58 | 15.36 | 2.26 |
| ITO/ o-MWCNTs | 3.4 | 47,933 | 2.33 | 22.73 | 2.42 |

**Table 1.** Whole device performance for both reference and target devices.

Figure 3b shows current efficiency-current density characteristics for reference and target devices. Compared with the reference device, the maximum current efficiency is greatly improved from 15.36 to 22.73 cd/A. As the current density increases, the efficiency for the target device is relatively stable to a wide range of currents, while it decreases much faster for the reference device. In addition, the target device reaches the best current efficiency at the luminance of 10,588 cd/m$^2$ with a voltage of 9.4 V, and the current density at this point is 146 mA/cm$^2$, while for the reference device, it appears at a luminance of 1,042 cd/m$^2$ with a voltage of 8.6 V and current density of 21 mA/cm$^2$. A high efficiency at a low luminance, for example a few candelas per meter squared, is of no pratical interest, although it may have relevance in understanding fundamental properties of materials or device characteristics [43]. But high efficiency at high luminance and low voltage should help ensure a stable and efficient OLED of practical significance. The increase in hole injection to TPD attributed to the o-MWCNTs layer increases the number of excitons at the TPD/Alq$_3$ interface. In Table 1, we list the optimized device performance parameters for both reference and target devices, and the performance improvements observed are very significant. The results also can be applied to other organic electronic devices that need high current density injection (or extraction).

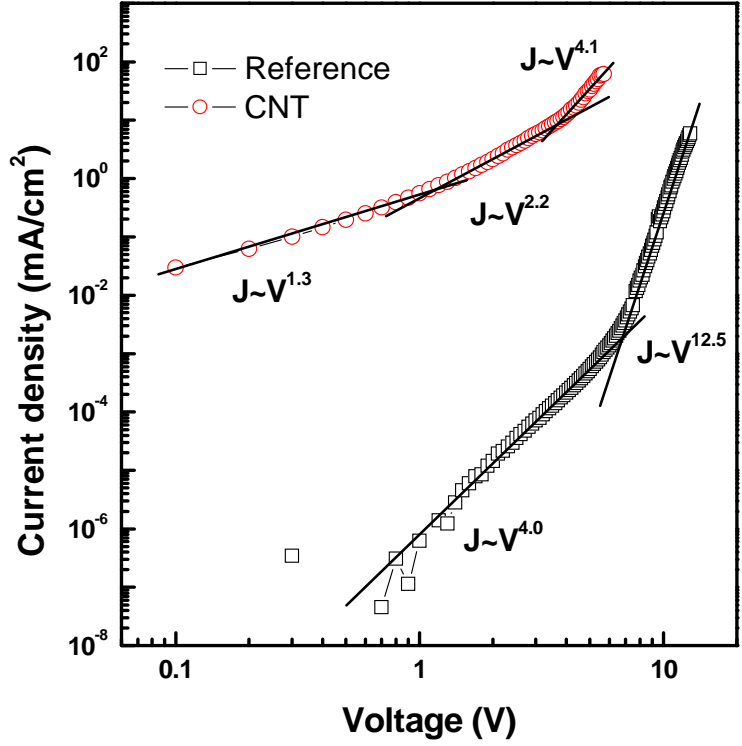

**Figure 4.** Current density-Voltage (J-V) characteristics for hole-only devices with pure (black square) and modified ITO (red circle) by o-MWCNTs on a log-log scale.

To elucidate the function of the o-MWCNT layer in the target device, we investigate the J-V characteristics for hole-only devices, with both pure and modified ITO (Figure 4). There is a current increase of more than $10^5$ times with the introduction of the o-MWCNT layer, especially at low voltages. It is clearly observed that the relationship of J–V obeys a power-law dependence, which has been proposed to be trap charge limited (TCL) [44]. This power-law dependence has been commonly observed in OLEDs. Based on the assumption of an exponential trap energy distribution, the J–V relationship can be described as [44]

$$J_{TCL} \propto V^{m+1}/d^{2m+1} \qquad (1)$$

where $m = T_t/T$ with T being the absolute temperature and $T_t$ being the characteristic temperature of the trap distribution, V is the applied voltage, and d is the active layer thickness. In Figure 4, the J-V relationship obeys an Ohmic conduction (J~V) at low voltage for the o-MWCNT hole-only device, and it becomes a space-charge-limited current (SCLC) model (J~V$^2$) as the voltage increases, and a higher component power-law with further increase of the voltage. While for the reference hole-only device, the current conduction is injection limited by contact effects, and it obeys power-law dependence with a large value of m, implying deep trap states at the interface. This difference can be described from the schematic diagrams of energy levels in the insert of Figure 2. Work function measurements on both pure and modified ITO were performed previously in our group by use of the Kelvin probe method, which measures the spatial average of the work function [37]. The effective work function for pure ITO is 4.7 eV, while it is increased to 5.0 eV for modified ITO with o-MWCNTs, with the TPD having a HOMO value of 5.4 eV. The high current density and SCLC behavior in the target hole-only device may be explained by the lower effective barrier height (~0.4 eV) for hole injection at anode/TPD interface, which is within the required threshold barrier height for the SCLC model (0.3-0.4 eV). While the low current density and higher exponent power-law dependence in the reference hole-only device can be explained by the higher effective barrier height (0.7 eV) for hole injection at the ITO/TPD interface, which results in an injection limited current.

Figure 5 shows AFM images of o-MWCNTs on top of ITO substrates on different scan sizes: 20 μm (a), 10 μm (b), and 5 μm (c), and the image of pure ITO substrate on the scale of 5 μm (d). The height scale is 20 nm for all the samples, and inserted arrows in Figure 5a-c indicate o-MWCNTs disperse on the ITO surface, and (b) and (c) are from the inserted dash rectangles in (a) and (b), respectively. The CNT quantity from Figure 5 (b) is estimated to be about 7 CNTs per 10 μm ×10 μm. AFM images show that the mean roughness increases slightly after the ITO modification by the low concentration o-MWCNTs. For example, at the same length scale, the average roughness for pure ITO is 2.8 nm in Fig 5(d), and that for the ITO modification by o-MWCNTs is 3.6 nm in Fig 5(c); and the o-MWCNT shapes can be observed in Figures 5a-c. Because of the low concentration (0.01 mg/ml), o-MWCNTs are well dispersed on the surface of the ITO substrate with a length of about 600 nm based on AFM images, and the diameter of o-MWCNT can be estimated to be about 12 nm from Fig 5(c).

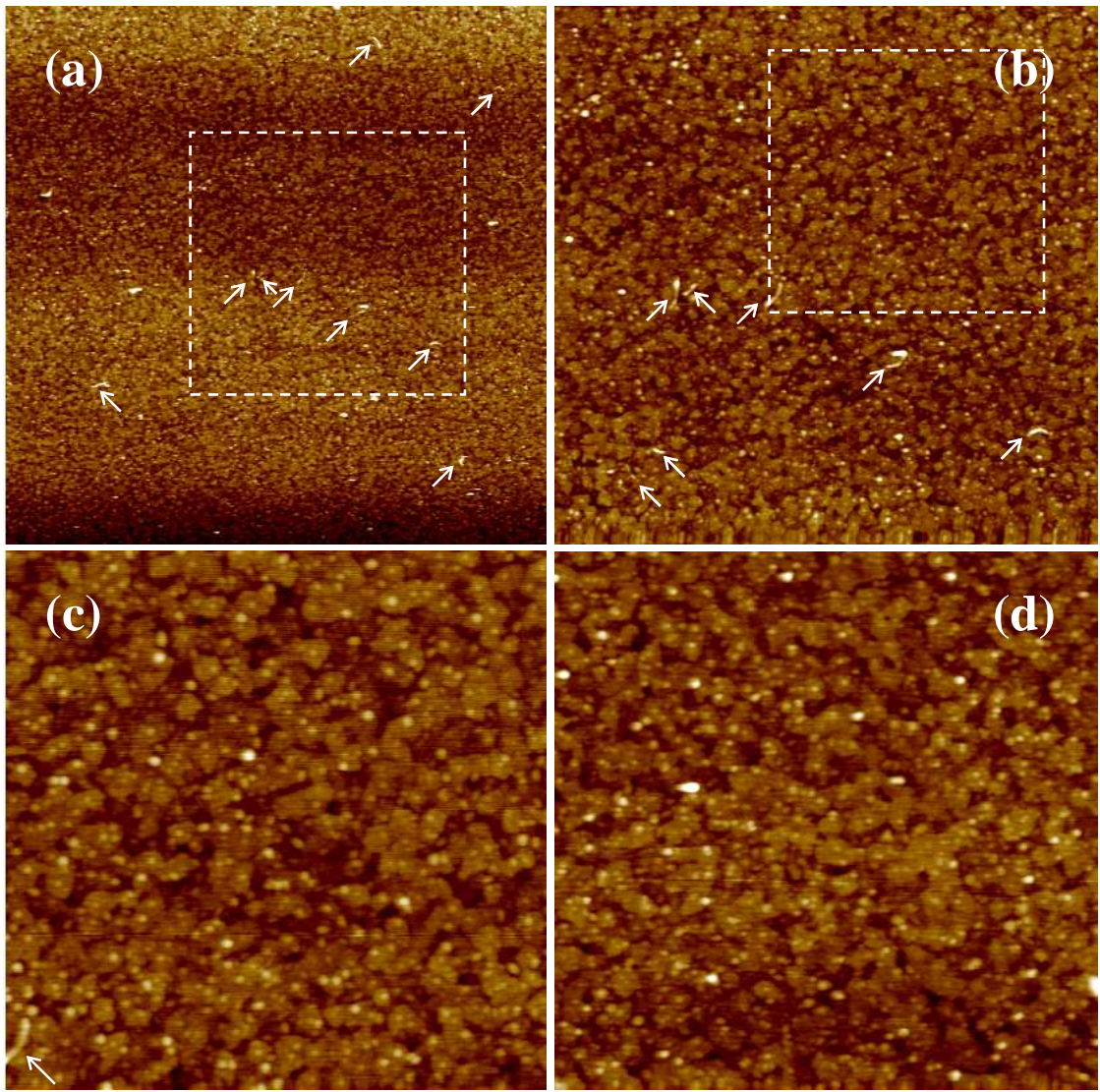

**Figure 5.** AFM images of o-MWCNTs on top of ITO substrates on different scan sizes: 20 μm **(a)**, 10 μm **(b)**, and 5 μm **(c)**, and the image of pure ITO substrate on the scale of 5 μm **(d)**. The height scale is 20 nm for all the samples. The inserted arrows in (a) to (c) indicate the o-MWCNTs disperse on ITO surface, and (b) and (c) are from the inserted dash rectangles in (a) and (b), respectively.

Small variations within the bulk of the nano-composite materials due to localised current channels, as found with CNTs, can affect the entire charge transport dynamics within the hybrid organic system [39,45]. Enhanced field-emission from CNT structures due to the large height to diameter aspect ratio of the structures is noted [46,47], and this can be replicated even when the structures are embedded within a polymer matrix such as a nanocomposite [48-50]. It has been shown that excellent electron emission can be obtained even when CNT concentrations as low as 0.7% volume fraction has been used in the composite, and small dispersions of MWCNTs will change the sheet resistance with even mono-layer coverage. Also very low concentration was reported to provide better and high quality dispersion of SWCNTs in solutions, as the concentration was decreased the number of individual SWCNTs increased from 20% to 50% [51]. In our case, the mean roughness is increased with the introduction of o-MWCNTs on ITO, and from Fig 5(a) and (b), we can find the aligned MWCNTs on the surface of ITO, but not straight. We can expect the enhancement of the local electric field within the plane of the o-MWCNTs, which improves hole injection from anode to organic active layer at much lower applied electric field. The electric filed enhancement can be due to both local geometry as well as differences in relative permittivity or dielectric inhomogeneity. The enhancement of the local electric field can be brought about by dielectric inhomogeneities between the matrix and the o-MWCNTs, and these inhomogeneities originate from the differences between conductive, spatially localized $sp^2$ C clusters surrounded by a more insulating $sp^3$ matrix [45]. Similar variations are observed in the propagation of electromagnetic waves in the case of plasmonics in metal sphere embedded dielectrics [47,52]. The discussion in the case of our composite films is based on the intrinsic properties of carbon nanotubes in comparison to its surroundings. Local geometry is probably another cause for local EL improvement, but we do not believe this to be the dominant process in this case.

Although hole injection is enhanced with the introduction of o-MWCNTs layer, there is no change in the EL spectra (not shown). The EL spectrum of the target device is identical to that of the reference device and is also independent of the drive voltage. Both devices have the same EL spectra around 532 nm, which means that the hole injection improvement doesn't alter the carrier recombination region. On the one hand, because of the enhancement of hole injection and its relatively high mobility, more holes can be transported into the $Alq_3$ layer to recombine with electrons injected from the cathode, but the thick layer of $Alq_3$ and buffer layer of LiF may restrict the holes from reaching the cathode and decrease the hole quenching prior to recombination. On the other hand, electron injection becomes easier because of the LiF buffer layer, but because the electrons have a much lower mobility than holes, the thick layer of $Alq_3$ will impede electron transport to the TPD layer and therefore keep electrons within the $Alq_3$ layer to some extent. The electron has too high a barrier (0.9 eV) to overcome to enter the TPD layer. Therefore, the carrier recombination zone is more or less fixed in the $Alq_3$ layer, near the interface of the TPD and $Alq_3$.

4. Conclusion

In summary, we have demonstrated the application of o-MWCNTs in OLEDs based on $Alq_3$ with a simple device structure, and we obtain very high luminance at relatively lower voltage which may be compared with the best related work on undoped $Alq_3$ fluorescent devices. As an efficient hole injection layer, o-MWCNTs improve the hole injection and allow for a better balanced hole and electron recombination mechanism. At only 10 V, the luminance can reach nearly 50,000 $cd/m^2$, with an external quantum efficiency over 2%, and a current efficiency greater than 21 cd/A. The results will unlock the route of applying CNTs in flexible OLEDs and other organic electronic devices. The technology we have developed is a generic one that may be applied to any high performance electrodes that need high current density injection (or extraction).

**Acknowledgements**

This research was partly funded by a portfolio partnership award by the EPSRC, UK, and by E.ON AG, as part of the E.ON International Research Initiative. Responsibility for the content of this publication lies with the authors.

**FIGURE CAPTIONS**

**Figure 1.** Transmittance for pure ITO (black) and modified ITO (red) by o-MWCNTs.

**Figure 2.** Current density-Voltage-Luminance (J-V-L) characteristics for OLEDs with pure (black square) and modified ITO (red circle) by o-MWCNTs. Inserted are the schematic diagrams for energy levels in the experiment for (top) reference device, and (down) target device with o-MWCNT modified ITO.

**Figure 3 (a).** External quantum efficiency (EQE)-Current density and **(b)** Current efficiency (CE)-Current density characteristics for OLEDs with pure (black square) and modified ITO (red circle) by o-MWCNT.

**Figure 4.** Current density-Voltage (J-V) characteristics for hole-only devices with pure (black square) and modified ITO (red circle) by o-MWCNTs on a log-log scale.

**Figure 5.** AFM images of o-MWCNTs on top of ITO substrates on different scan sizes: 20 μm **(a)**, 10 μm **(b)**, and 5 μm **(c)**, and image of pure ITO substrate on the scale of 5 μm **(d)**. The height scale is 20 nm for all the samples. The inserted arrows in (a) to (c) indicate the o-MWCNTs disperse on ITO surface, and (b) and (c) are from the inserted dash rectangles in (a) and (b), respectively.

**Table 1.** Whole device performance for both reference and target devices